# Measuring collaborative emergent behavior in multi-agent reinforcement learning

Sean L. Barton[1], Nicholas R. Waytowich[2], Erin Zaroukian[1], & Derrik E. Asher[1]

[1]Computational & Information Sciences Directorate, U.S. Army Research Laboratory
[2]Human Research & Engineering Directorate, U.S. Army Research Laboratory

**Abstract.** Multi-agent reinforcement learning (RL) has important implications for the future of human-agent teaming. We show that improved performance with multi-agent RL is not a guarantee of the collaborative behavior thought to be important for solving multi-agent tasks. To address this, we present a novel approach for quantitatively assessing collaboration in continuous spatial tasks with multi-agent RL. Such a metric is useful for measuring collaboration between computational agents and may serve as a training signal for collaboration in future RL paradigms involving humans.

## 1      Introduction

Reinforcement learning (RL) is an attractive option for providing adaptive behavior in computational agents because of its theoretical generalizability to complex problem spaces [1, 2]. In particular, deep RL recently produced striking results [3]. Extending RL to the multi-agent domain has received an increasing amount of attention as the need for human-agent teams has increased, especially with regards to training agents to behave collaboratively [1, 4–8].

Unfortunately, the nature of multi-agent RL makes guaranteeing collaboration between agents impossible except in limited provable cases, even when these methods yield better task performance [1, 7]. Thus far, evaluating collaboration in multi-agent learning has been accomplished by measuring performance in tasks where coordination is required. While this may be satisfactory for discretized tasks where cooperative policies are provably optimal [1, 6–8] it is not clear that this methodology generalizes well to more complex and continuous tasks (such as those presented in [4, 5] and here).

The question at hand is how to assess performance enhancing coordination (or collaboration) between agents in multi-agent RL tasks. Here, we present a method borrowed from the field of ecology, called convergence cross mapping (CCM), and show how it can be used to measure collaboration between agents during a predator-prey pursuit task. Additionally, we show a striking result: a state-of-the-art multi-agent reinforcement learning algorithm does not exhibit coordinated behavior between agents during a collaborative task even though high task performance is achieved, indicating that performance metrics alone are not sufficient for measuring collaboration.

## 2 Methods

### 2.1 Simulation Environment

In a modification of the classic predator-prey pursuit task (see Figure 1A-C), three slower predator agents score points each time they make contact with a prey agent in a continuous bounded 2D particle environment. Predator agents were identical in terms of capabilities (i.e., velocity and acceleration). This simulation environment was made available through the OpenAI Gym network [9] and was developed for the multi-agent deep learning algorithm discussed below [4].

Prey agents were capable of 33% greater acceleration and 25% greater maximum velocity than any predator agent, making capture by a solitary predator extremely difficult. All agents had the same mass, with minimal elastic properties to provide a small bump force upon collisions. Agent positions and velocities were randomized at the start of each episode, and acceleration was initially set to zero. Predator agents all received a fixed reward when any one of them made contact with the prey agent. Prey agents received a punishment when they were contacted.

### 2.2 Agents

In order to evaluate our metric's ability to estimate collaboration between predators in the predator-prey task, we utilized two types of agents: learning agents and fixed-strategy (non-learning) agents. Learning agents' behaviors were guided by a multi-agent deep deterministic policy gradient (MADDPG) algorithm [4]. In all cases, prey agents were learning agents and thus utilized the MADDPG algorithm independent of predator behavior.

Two types of distinct fixed-strategy predators were implemented to demonstrate upper and lower bounds of coordinated behaviors. The first (termed 'Chaser' predators, see Figure 1A) naively pursued prey agents by maximally accelerating in the instantaneous direction of the prey relative to the predator's own position. As such, Chasers were incapable of coordinating their behavior with each other.

The second strategy (termed 'Spring' predators, see Figure 1B) also naively minimized distance to the prey, but predator movements were modified by spring forces which constrained their position and velocity relative to one another. A Spring predator's movement direction was a sum of the spring forces acting on it and its desired vector of movement towards the prey. In this case, predator actions were explicitly coordinated with their partners'.

The MADDPG algorithm (Figure 1C and 1E) used to guide learning agent behavior is an extension of a deep deterministic policy gradient (DDPG) algorithm [3] into the multi-agent domain [4]. Like DDPG, MADDPG utilizes an actor-critic model with deep neural networks representing policy and Q-learners (Figure 1E). Multi-agent capabilities are achieved by passing information about each agent's state and actions to each critic network. As such, the learning agents are *joint action learners* as opposed to *independent learners*, giving them an advantage when coordinating their behaviors [8].

All conditions (Figure 1A-C) were trained for 100k episodes to ensure model convergence before evaluating collaborative behavior (Figure 1D). Each episode lasted for 25 time steps during learning and 2000 time steps during evaluation. These intervals were selected to match what was shown in literature for sufficient learning [4] and required for analysis [10]. During testing, 10 episodes were recorded without learning for each predator strategy in order to produce a distribution of agent behaviors with the same level of training over various random starting positions.

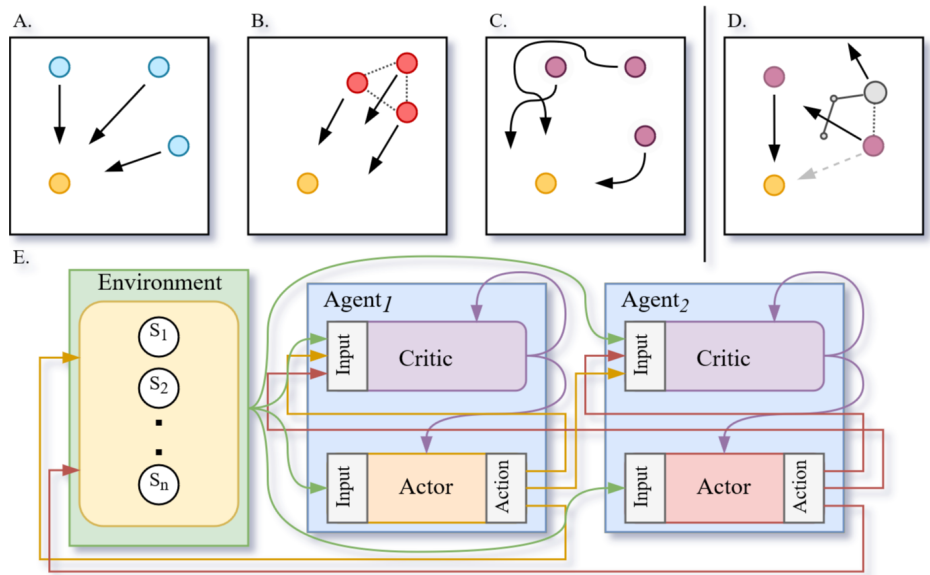

**Fig. 1.** Experimental design. A-C) Yellow circles denote prey, while others represent predator strategies. Black arrows indicate actions. Panels show A) Chaser, B) Spring (coupling shown as grey dotted lines), and C) MADDPG agent conditions. D) At test time, one predator's behavior was determined by the physics of a double pendulum (grey circle). Agents coordinated with a modified predator should have their goal-seeking actions (grey dashed arrow) augmented by the actions of the modified predator. E) Schematic of MADDPG algorithm.

### 2.3 Collaboration Metrics

In order to measure collaboration, we used predators' positions over time as time-series data to implement a technique called convergent cross mapping (CCM) which examines the causal influence one time-series has on another [10]. The CCM technique embeds a time-series in a high-dimensional attractor space and then uses this embedded data as a model to predict states of the other time-series in its own attractor space. To the extent that this is possible, the original time-series is said to be causally driven by the time-series it attempts to model. Thus, in multi-agent tasks with homogeneous agents, we are defining collaboration to be the amount of causal influence between agents as measured by the CCM.

Importantly, this metric can be misleading when two time-series have a mutually causal relationship with a third. For the present task, this is important because the causal influence one predator has on another can be confounded by their mutual relationship with the prey. To address this, at test time we modified the behavior of one predator agent such that its previous behavior was replaced by a secondary behavior that did not pursue the prey[1]. If a causal influence exists between predators, the movements of the modified predator should change the behavior of the other predators, even though the modified predator no longer pursues the prey (see Figure 1D).

## 3  Results

Table 1 shows the log scaled mean reward per episode for predator and prey agents in the Chaser, Spring, and MADDG cases. It is clear that MADDPG-equipped predators are betters performers of this pursuit task, achieving roughly an order of magnitude more average reward per trial.

**Table 1**. Performance results for the different experimental conditions

|         | N  | Mean Reward | SD   | 95% CI |
|---------|----|-------------|------|--------|
| Chaser  | 10 | 1.85        | 0.66 | 0.47   |
| Spring  | 10 | 4.29        | 0.92 | 0.66   |
| MADDPG  | 10 | 13.70       | 2.57 | 1.84   |

While the capacity for coordinated actions afforded by the MADDPG algorithm coupled with the increased task performance may suggest collaboration between predator agents, the CCM analysis tells a different story (Figure 2). As shown, there is very little difference, in terms of causal influence, between MADDPG-equipped predators and naive Chasers. The causal influence of the modified agent might be marginally stronger for MADDPG predators (hinting at the shared action information afforded by this algorithm), but the effect is too weak to make any strong conclusions about the collaboration between the agents in this case.

For Spring predators, on the other hand, there is a strong causal influence from the modified predator. The Spring experimental condition (see Figure 2) shows an example of coupled behaviors in the predator-prey pursuit task. This provides a strong validation for the CCM technique, and illustrates an ecological upper-bound on coordinated actions between learning agents.

---

[1] In this case, the dynamics of a double pendulum were used to specify the movement of the modified predator.

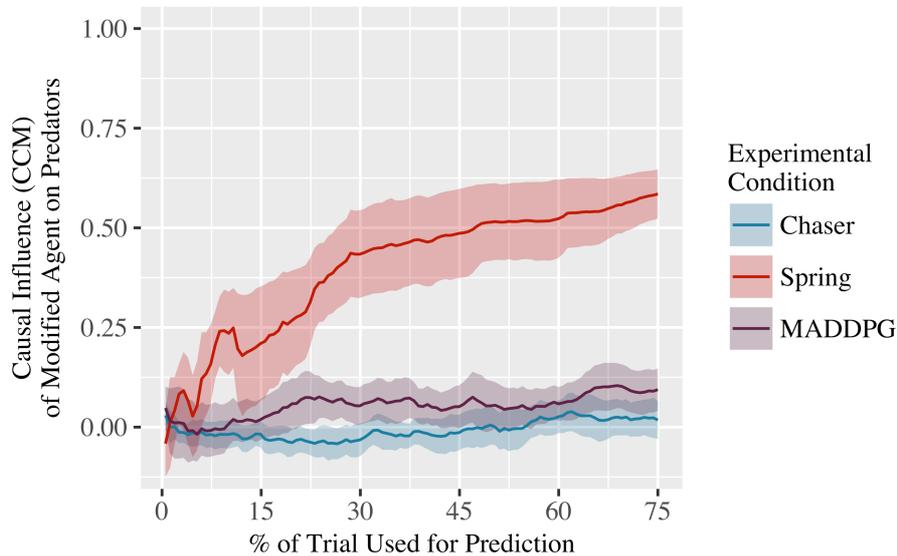

**Fig. 2.** CCM measure for causal influence of modified predator behavior on unmodified predators. True causal influence is indicated by an increase in CCM score (Y axis) as more of a trial (X axis) is used to build the model (so called "convergence" [10]). A clear causal relationship is shown for Spring predators. Minimal causal influence is detectable for MADDPG agents. Coordination for Chaser agents is indistinguishable from zero.

## 4 Discussion

The MADDPG algorithm presented by Lowe et al. [4] is promising, because it suggests that multi-agent deep RL may be able to solve complex problems in continuous tasks involving multiple actors. However, as we demonstrate here the improved performance of the MADDPG algorithm does not necessarily indicate coordinated behavior between agents.

Assessing collaboration between agents instead requires a direct measure of the coordination between agent actions. We present such a method here, in the form of CCM, and we validate the metric within the context of a continuous multi-agent task. Though we did not find strong evidence of coordination between the actions of learning predators, the marginal increase in coordination over Chaser predators points to potential for collaboration given sufficient learning pressure.

Producing collaborative behavior in computational agents is critical for the future of human-agent teams. Human-factors research has long held that when computational systems fail to adapt to human needs, serious issues in performance and human satisfaction can arise [11]. These issues are often best alleviated by promoting a collaborative relationship between humans and computational agents, rather than forcing humans to act as overseers [12, 13]. Measures like CCM, which can be used to assess (and even promote) collaborative behaviors in multi-agent RL, constitute powerful tools for the future of human-computer interactions.

**Acknowledgements and Disclosure:** This research was sponsored by the Army Research Laboratory and was accomplished under Cooperative Agreement Number W911NF-18-2-0058. The views and conclusions contained in this document are those of the authors and should not be interpreted as representing the official policies, either expressed or implied, of the Army Research Laboratory or the U.S. Government. The U.S. Government is authorized to reproduce and distribute reprints for Government purposes notwithstanding any copyright notation herein.

# References

1. Matignon, L., Laurent, G.J., Le Fort-Piat, N.: Independent reinforcement learners in cooperative markov games: A survey regarding coordination problems. The Knowledge Engineering Review. 27, 1–31 (2012).
2. Sen, S., Sekaran, M., Hale, J., others: Learning to coordinate without sharing information. In: AAAI. pp. 426–431 (1994).
3. Mnih, V., Kavukcuoglu, K., Silver, D., Rusu, A.A., Veness, J., Bellemare, M.G., Graves, A., Riedmiller, M., Fidjeland, A.K., Ostrovski, G., Petersen, S., Beattie, C., Sadik, A., Antonoglou, I., King, H., Kumaran, D., Wierstra, D., Legg, S., Hassabis, D.: Human-level control through deep reinforcement learning. Nature. 518, 529–533 (2015).
4. Lowe, R., WU, Y., Tamar, A., Harb, J., Pieter Abbeel, O., Mordatch, I.: Multi-Agent Actor-Critic for Mixed Cooperative-Competitive Environments. In: Guyon, I., Luxburg, U.V., Bengio, S., Wallach, H., Fergus, R., Vishwanathan, S., and Garnett, R. (eds.) Advances in Neural Information Processing Systems 30. pp. 6382–6393. Curran Associates, Inc. (2017).
5. Foerster, J., Farquhar, G., Afouras, T., Nardelli, N., Whiteson, S.: Counterfactual Multi-Agent Policy Gradients. arXiv:1705.08926 [cs]. (2017).
6. Matignon, L., Laurent, G., Le Fort-Piat, N.: Hysteretic q-learning: An algorithm for decentralized reinforcement learning in cooperative multi-agent teams. In: IEEE/rsj international conference on intelligent robots and systems, iros'07. pp. 64–69 (2007).
7. Lauer, M., Riedmiller, M.: An algorithm for distributed reinforcement learning in cooperative multi-agent systems. In: In proceedings of the seventeenth international conference on machine learning. Citeseer (2000).
8. Claus, C., Boutilier, C.: The dynamics of reinforcement learning in cooperative multiagent systems. AAAI/IAAI. 1998, 746–752 (1998).
9. Brockman, G., Cheung, V., Pettersson, L., Schneider, J., Schulman, J., Tang, J., Zaremba, W.: OpenAI Gym. arXiv:1606.01540 [cs]. (2016).
10. Sugihara, G., May, R., Ye, H., Hsieh, C.-h., Deyle, E., Fogarty, M., Munch, S.: Detecting causality in complex ecosystems. science. 1227079 (2012).
11. Parasuraman, R., Sheriden, T.B., Wickens, C.D.: A model for types and levels of human interaction with automation. IEEE Transactions on Systems, Man, and Cybernetics - Part A: Systems and Humans. 30, 286–297 (2000).
12. Rovira, E., McGarry, K., Parasuraman, R.: Effects of Imperfect Automation on Decision Making in a Simulated Command and Control Task. Human Factors. 49, 76–87 (2007).
13. Klein, G., Woods, D.D., Bradshaw, J.M., Hoffman, R.R., Feltovich, P.J.: Ten challenges for making automation a "team player" in joint human-agent activity. IEEE Intelligent Systems. 19, 91–95 (2004).